\documentclass[12pt]{article}
\usepackage{amsmath}
\usepackage{amssymb}
\def\Stran       {\mathcal{S}}
\def\Ttran       {\mathcal{T}}
\def\N           {\mathbb N}
\def\Z           {\mathbb Z}
\def\R           {\mathbb R}
\def\C           {\mathbb C}
\def\defd        {\stackrel{\rm d}{=}}
\def\su          {\hat{su}(2)}
\def\isok        {\hat{sl}(2;\mathbb C)_k}

\def\hslc        {\hat{sl}(2|1;{\mathbb C})}
\def\hslck       {\hat{sl}(2|1;{\mathbb C})_k}
\def\hslcp       {\hat{sl}(2|1;{\mathbb C})_{-\frac{1}{2}}}
\def\hslcq       {\hat{sl}(2|1;{\mathbb C})_{-\frac{2}{3}}}
\def\hf          {\frac{1}{2}}
\def\thf         {\frac{3}{2}}
\def\egt         {\frac{1}{8}}
\def\thrd        {\frac{1}{3}}
\def\tthrd       {\frac{2}{3}}
\def\fthrd       {\frac{4}{3}}
\def\vthrd       {\frac{5}{3}}
\def\sxth        {\frac{1}{6}}
\def\rui       {e^{\pi i/3}}
\def\rumi      {e^{-\pi i/3}}
\def\ruii      {e^{2\pi i/3}}
\def\rumii     {e^{-2\pi i/3}}
\begin{document}

\numberwithin{equation}{section}

\begin{titlepage}
\begin{flushright}
DTP/99/59\\
December 1999\\
\end{flushright}
\vspace{1cm}
\begin{center}
{\Large\bf Modular Transformations and Invariants in the Context of Fractional Level $\hslc$}\\
\vspace{1cm}
{\large Gavin Johnstone}\\
\vspace{0.5cm}
{\it Department of Mathematical Sciences, University of Durham, \\Durham, DH1 3LE, England}\\
\vspace{0.5cm}
e-mail: G.B.Johnstone@durham.ac.uk
\end{center}
\vspace{2cm}

\begin{abstract}

The modular transformation properties of admissible characters of the affine superalgebra $\hslck$ at fractional level $k=1/u-1$, $u \in \N \setminus \{1\}$ are presented.  All modular invariants for $u=2$ and $u=3$ are calculated explicitly and an $A$-series and $D$-series of modular invariants emerge.

\end{abstract}

\end{titlepage}

\section{Introduction}

In a series of papers \cite{BT97, BHT98, HT98}, the properties of the affine superalgebra $\hslck$ at fractional level were extensively investigated.  As well as the abstract interest, the motivation for this work was in the potential relevance to the $N=2$ non-critical string.  The generally held notion that a non-critical (super)string is described in terms of a topological $G/G$ Wess-Zumino-Novikov-Witten model, with $G$ a Lie (super)group \cite{Y, FY, HY}, provides the link: in order to describe the spectrum in the $G=SL(2|1;\R)$ case, believed to be that related to the $N=2$ string, a good understanding of $\hslck$ at fractional level is required.  Indeed, when  the matter, which is coupled to supergravity in the $N=2$ non-critical string, is minimal, $i.e.$ taken in an $N=2$ super Coulomb gas representation with central charge
\begin{equation}
c_{\text{matter}}=3\left(1-\frac{2p}{u}\right), \qquad p, u \in \N, \qquad {\rm gcd}(p,u)=1,
\end{equation}
it can be shown that the level of the ``matter'' affine superalgebra $\hslck$ appearing in the $SL(2|1;\R)/SL(2|1;\R)$ model is of the form
\begin{equation}
k = \frac{p}{u} - 1.
\end{equation}
It is precisely these values of the level for which admissible representations of $\hslck$ do exist \cite{KW88}.

The present work builds immediately on that of \cite{HT98}, inheriting this motivation, but also with pretensions to more general relevance in that we construct modular invariant combinations of $\hslck$ characters, which could be taken as starting points in the building of partition functions for rational conformal field theories.  

We begin in section 2 by presenting the branching functions of so-called class $IV$ and class $V$ $\hslck$ characters\footnote{Class $I$ characters are modular forms rather than functions and it is not clear if they can be used to construct modular invariants.  Class $II$ and class $III$ representations are believed to contain subsingular vectors, making the computation of their characters extremely difficult; by contrast, classes $IV$ and $V$ are believed to contain no subsingular vectors, making the calculation of their characters tractable.  They are also modular functions, justifying their r\^ole as the central objects of study.  Incidentally, those which are singular in the limit $\sigma \rightarrow 0$ are related to $N=2$ superconformal characters.  See \cite{BT97}, \cite{BHT98} and \cite{HT98} for definitions and detailed exposition.}, where the level $k=1/u-1$, $u \in \N \setminus \{1\}$.  These branching functions provide expressions for $\hslck$ characters in terms of functions with known modular transformation properties, facilitating the analysis of their behaviour under modular transformations.  

The branching functions were first calculated (actually conjectured, based upon analysis for low values of $u$) in \cite{HT98} for Neveu-Schwarz characters (corresponding to the twisted version of $\hslck$).  This was a natural starting point since the challenge here lies in computing the effect of the modular $\Stran$ transformation $\tau \rightarrow -\frac{1}{\tau}$, under which Neveu-Schwarz characters transform into a linear combination of Neveu-Schwarz characters at the same level.  However, we will go on to consider the question of forming modular invariant combinations of characters: as the modular $\Ttran$ transformation $\tau \rightarrow \tau +1$ takes Neveu-Schwarz characters to supercharacters, which under $\Stran$ transform into Ramond characters (corresponding to the untwisted version of $\hslck$), we will require knowledge of the modular transformation properties of these characters as well.  To this end, we calculate Neveu-Schwarz supercharacter and Ramond character branching functions, simply derived by spectral flow.  Under $\Ttran$, Ramond characters do give back a combination of Ramond characters so consideration of this set is enough to allow us to form modular invariants; however, for completeness we also consider Ramond supercharacters, which form a closed set under modular transformations.

Having established branched expressions for the various $\hslck$ characters such that they are amenable to applying the modular $\Stran$ transformation, we proceed to do so in section 3.  Here we derive the action of the $\Stran$ transformation on the $\hslck$ Neveu-Schwarz and Ramond characters and supercharacters at all levels $k=1/u-1$, $u \in \N \setminus \{1\}$.  We also mention the action of the modular $\Ttran$ transformation.   

Finally in section 4 we identify an $A$-series and $D$-series of modular invariants, based on the appearance of such invariants in the two simplest cases of $u=2$ and $u=3$, the details of which may be found in the appendices.

\section{Branching $\hslck$ Characters}

In \cite{HT98}, branching formulae for the Neveu-Schwarz class $IV$ and class $V$ characters of the affine superalgebra $\hslck$ ($k=1/u-1$) were conjectured.  The characters were branched into products of $\isok$ characters, generalised theta functions and string functions, the modular transformation properties of which are known.

 By definition, $\hslck$ characters are given by
\begin{equation}
\chi^{\hslck}_{h_-,h_+}
(\sigma,\nu,\tau)\defd\text{tr}\exp \{2\pi i (\sigma
J_0^3+\nu U_0+\tau L_0)\},
\end{equation}
where $J_0^3$ and $U_0$ are the zero mode Cartan generators of $\hslck$.  We label the characters by the isospin and charge quantum numbers which characterise the $\hslck$ highest weight states $|\Omega \rangle$ of the associated representations\footnote{See \cite{BT97, BHT98} for more details.  It is conditions on these quantum numbers that split the representations into the classes mentioned earlier.}:
\begin{equation}
J_0^3 |\Omega \rangle = \hf h_- |\Omega \rangle, \quad U_0 |\Omega \rangle=\hf h_+ |\Omega \rangle.
\end{equation}

From ref. \cite{MP90} we have the following expression for $\isok$ characters:
\begin{equation}
\label{sl2th}
\chi^{\isok}_{n,n'}(\sigma,\tau) = 
                \frac{\vartheta_{v_+,w}\left(\frac{\sigma}{u}, \tau \right)
                - \vartheta_{v_-,w}\left(\frac{\sigma}{u}, \tau \right)}
        {\vartheta_{1,2}(\sigma, \tau) - \vartheta_{-1,2}(\sigma, \tau)},
\end{equation}
where the level is parametrised as
\begin{equation}
k=\frac{t}{u},\qquad \text{gcd}(t,u)=1,\quad u \in \N,\quad t \in \Z,
\end{equation}
with $0\leqslant n\leqslant 2u+t-2 \text{ and } 0\leqslant n'\leqslant u-1$ and
\begin{equation}
\label{vw}
v_{\pm}\defd u(\pm(n+1) - n'(k+2)),\qquad w\defd u^2(k+2).
\end{equation}
In the above, the generalised theta functions $\vartheta_{m,m'}$ \cite{Kbook} are defined as 
\begin{equation}
\label{thfn}
\vartheta_{m,m'}(\sigma, \tau) \defd \sum_{n\in\Z}q^{m'(n +\frac{m}{2m'})^2}z^{m'(n + \frac{m}{2m'})}. 
\end{equation}
The variables $q$ and $z$ are defined by
\begin{align}
q &\defd \text{exp}(2\pi i\tau), \quad \tau\in\C,\; \text{Im}(\tau) > 0 \Rightarrow |q| < 1,\notag\\
z &\defd \text{exp}(2\pi i\sigma).
\end{align}
We are interested in the cases where $k=1/u-1$, that is, where $t=1-u$.

Under  the modular $\Stran$ transformation $\Stran : (\sigma, \nu, \tau) \rightarrow \left(\frac{\sigma}{\tau}, \frac{\nu}{\tau}, -\frac{1}{\tau}\right)$, the $\isok$ characters \eqref{sl2th} transform via \cite{MP90}
\begin{equation}
\label{isotr}
\chi ^{\isok}_{m,m'}\left(\frac{\sigma}{\tau},-\frac{1}{\tau}\right) =
e^{-i\pi k \sigma^2/\tau}\sum_{n=0}^{2u+t-2} \sum_{n'=0}^{u-1}
S_{mm', nn'} \chi^{\isok}_{n, n'}(\sigma,\tau),
\end{equation}
where
\begin{equation}
S_{mm',nn'}=\sqrt{\frac{2}{u^2(k+2)}}(-1)^{m'(n +1)+(m+1)n'} e^{-i\pi (k+2)m'n'} \sin \left(\frac{\pi (m+1)(n +1)}{k+2}\right).
\end{equation}

For the generalised theta functions \eqref{thfn} we have \cite{Kbook}
\begin{equation}
\label{thetatr}
\vartheta_{m, m'}\left(\frac{\sigma}{\tau}, -\frac{1}{\tau}\right) = 
e^{-i\pi m' \sigma^2/\tau} \sqrt{\frac{-i\tau}{2m'}}\,
\sum_{r=0}^{2m'-1} e^{-i\pi rm/m'}\vartheta_{r,m'}(\sigma, \tau)
\end{equation}
and for the string functions \cite{GQ87}
\begin{equation}
\label{ctr}
c^{(u-1)}_{a,b}\left(-\frac{1}{\tau}\right)=\frac{1}{\sqrt{(-i\tau)(u-1)(u+1)}}\,\underset{a'\equiv b' \; \text{mod }2}{\sum^{u-1}_{a'=0}\,\sum^{u-1}_{b'=-u+2}}s(a,b,a',b')\,c^{(u-1)}_{a',b'}(\tau),
\end{equation}
where
\begin{equation}
s(a,b,a',b')=e^{i\pi bb'/(u-1)}\,\text{sin}\left(\frac{\pi(a+1)(a'+1)}{u+1}\right).
\end{equation}

The string functions have the following symmetries \cite{KP}:
\begin{align}
\label{csym}
c^{(u-1)}_{a,b}(\tau)=c^{(u-1)}_{a,-b}(\tau) &=
c^{(u-1)}_{a,b+2(u-1)\Z}(\tau)=c^{(u-1)}_{u-1-a,u-1-b}(\tau),\notag\\
c^{(u-1)}_{a,b}(\tau) &=0 \quad \text{for} \quad a-b \neq 0 \mod{2}.
\end{align} 

We now state the formulae for the branching of $\hslck$ Neveu-Schwarz characters as conjectured in \cite{HT98}.  The result for class $IV$ characters reads
\begin{multline}
\label{NSIVbr}
\chi_{h^{NS}_-,h^{NS}_+}^{NS,IV,\hslck}(\sigma, \nu, \tau)=
\sum_{a=0}^{u-1}  \chi^{\isok}_{a,u-m-1}(\sigma, \tau)\\
\times \sum _{b=0}^{u-2} 
\vartheta_{(u-1)(m-2m')+u(u-1)(a+1)+2au(\frac{u}{2}-[\frac{u}{2}])-2ub,u(u-1)} \left(\frac{\nu}{u}, \tau\right)\\
\times c^{(u-1)}_{a, a(u-1)+2a(\frac{u}{2}-[\frac{u}{2}])-2b}(\tau)
\end{multline}
and for those in class $V$ it is
\begin{multline}
\label{NSVbr}
\chi_{h^{NS}_-,h^{NS}_+}^{NS,V,\hslck}(\sigma, \nu, \tau)=
\sum_{a=0}^{u-1} \chi^{\isok}_{a,M+M'+1}(\sigma,\tau) \\
\times \sum _{b=0}^{u-2}
\vartheta_{(u-1)(M'-M)+u(u-1)a+2au(\frac{u}{2}-[\frac{u}{2}])-2ub,u(u-1)} \left(\frac{\nu}{u}, \tau\right) \\
\times c^{(u-1)}_{a, a(u-1)+2a(\frac{u}{2}-[\frac{u}{2}])-2b}(\tau),
\end{multline}
where $[\frac{u}{2}]$ denotes the integer part of $\frac{u}{2}$.
  
For the Neveu-Schwarz sector, the isospin ($\hf h^{NS}_-$) and charge ($\hf h^{NS}_+$) quantum numbers are given by
\begin{equation}
h_-^{NS}=-\frac{1}{u}(u-m-1),\quad h_+^{NS}=\frac{1}{u}(2m'-m),\quad 0 \leqslant m' \leqslant m \leqslant u-1
\end{equation}
in class $IV$ and by
\begin{equation}
h_-^{NS}=-\frac{1}{u}(M+M'+1+u),\quad h_+^{NS}=\frac{1}{u}(M-M'), \quad 0 \leqslant M+M' \leqslant u-2
\end{equation}
in class $V$. 
 
As we will come to consider the characters and supercharacters of the Ramond sector and the supercharacters of the Neveu-Schwarz sector, we mention that the Neveu-Schwarz supercharacters have the same quantum numbers as the characters while the Ramond characters and supercharacters have
\begin{equation}
h_-^R=-\frac{m}{u},\quad h_+^R=h_+^{NS},\quad 0 \leqslant m' \leqslant m \leqslant u-1
\end{equation}
in class $IV$ and
\begin{equation}
h_-^R=\frac{1}{u}(M+M'+2),\quad h_+^R=h_+^{NS}, \quad 0 \leqslant M+M' \leqslant u-2
\end{equation}
in class $V$.  The conformal weight in both classes is given by
\begin{equation}
h^R=\frac{u}{4}\left((h_-^R)^2-(h_+^R)^2\right) 
\end{equation}
and in the Neveu-Schwarz sector by
\begin{equation}
h^{NS}=h^R-\hf h_-^R+\frac{1-u}{4u}. 
\end{equation}

We can use spectral flow \cite{BHT98} to obtain the branching formulae for Ramond characters from \eqref{NSIVbr} and \eqref{NSVbr}:
\begin{equation}
\chi^{NS,\hslck}_{h_-^{NS},h_+^{NS}}(\sigma,\nu,\tau)=
q^{k/4}z^{ k/2}\chi^{R,\hslck}_{h_-^R,h_+^R}(-\sigma-\tau,\nu,\tau),
\end{equation}
giving
\begin{equation}
\chi^{R,\hslck}_{h_-^{R},h_+^{R}}(\sigma,\nu,\tau)=
q^{(1-u)/4u}z^{(1-u)/2u}\chi^{NS,\hslck}_{h_-^{NS},h_+^{NS}}(-\sigma-\tau,\nu,\tau)
\end{equation}
recalling that we are considering only $k=1/u-1$.  

Using the definition of the generalised theta functions \eqref{thfn} and the definition of the $\isok$ characters in terms of these given by \eqref{sl2th}, we find 

\begin{equation}
\chi^{\isok}_{n,n'}(-\sigma-\tau,\tau)=q^{-(1-u)/4u} z^{-(1-u)/2u}\chi^{\isok}_{u-1-n,u-1-n'}(\sigma,\tau)
\end{equation}
and thus
\begin{multline}
\label{RIVbr}
\chi_{h^{R}_-,h^{R}_+}^{R,IV,\hslck}(\sigma, \nu, \tau)=
\sum_{a=0}^{u-1}  \chi^{\isok}_{u-1-a,m}(\sigma,\tau)\\
\times \sum _{b=0}^{u-2} 
\vartheta_{(u-1)(m-2m')+u(u-1)(a+1)+2au(\frac{u}{2}-[\frac{u}{2}])-2ub,u(u-1)} \left(\frac{\nu}{u}, \tau\right) \\
\times c^{(u-1)}_{a, a(u-1)+2a(\frac{u}{2}-[\frac{u}{2}])-2b}(\tau)
\end{multline}
and in class $V$ 
\begin{multline}
\label{RVbr}
\chi_{h^{R}_-,h^{R}_+}^{R,V,\hslck}(\sigma, \nu, \tau)=
\sum_{a=0}^{u-1} \chi^{\isok}_{u-1-a,u-M-M'-2}(\sigma,\tau) \\
\times \sum _{b=0}^{u-2}
\vartheta_{(u-1)(M'-M)+u(u-1)a+2au(\frac{u}{2}-[\frac{u}{2}])-2ub,u(u-1)} \left(\frac{\nu}{u}, \tau\right) \\
\times c^{(u-1)}_{a, a(u-1)+2a(\frac{u}{2}-[\frac{u}{2}])-2b}(\tau),
\end{multline}
with definitions as given previously.
 
To obtain the Neveu-Schwarz supercharacters, we must shift the variable $\sigma\rightarrow \sigma +1$ in \eqref{NSIVbr} and \eqref{NSVbr}, then divide the results by $e^{i\pi h^{NS}_-}$ (this procedure corresponding to an insertion of the operator $(-1)^F$ in the definition of the character).  For class $IV$ we find
\begin{multline}
\label{SNSIVbr}
S\chi_{h^{NS}_-,h^{NS}_+}^{NS,IV,\hslck}(\sigma, \nu, \tau)=
\sum_{a=0}^{u-1} e^{i\pi(a-(u-m-1))} \chi^{\isok}_{a,u-m-1}(\sigma,\tau) \\
\times \sum _{b=0}^{u-2}
\vartheta_{(u-1)(m-2m')+u(u-1)(a+1)+2au(\frac{u}{2}-[\frac{u}{2}])-2ub,u(u-1)} \left(\frac{\nu}{u}, \tau\right)\\
\times c^{(u-1)}_{a, a(u-1)+2a(\frac{u}{2}-[\frac{u}{2}])-2b}(\tau)
\end{multline}
and for class $V$ we obtain
\begin{multline}
\label{SNSVbr}
S\chi_{h^{NS}_-,h^{NS}_+}^{NS,V,\hslck}(\sigma, \nu, \tau)=
\sum_{a=0}^{u-1} e^{i\pi (a-(M+M'))} \chi^{\isok}_{a,M+M'+1}(\sigma,\tau) \\
\times \sum _{b=0}^{u-2}
\vartheta_{(u-1)(M'-M)+u(u-1)a+2au(\frac{u}{2}-[\frac{u}{2}])-2ub,u(u-1)} \left(\frac{\nu}{u}, \tau\right)\\
\times c^{(u-1)}_{a, a(u-1)+2a(\frac{u}{2}-[\frac{u}{2}])-2b}(\tau).
\end{multline}

The Ramond sector supercharacters are obtained in similar fashion from \eqref{RIVbr} and \eqref{RVbr}, shifting $\sigma\rightarrow \sigma +1$ and dividing by $e^{i\pi h^{R}_-}$:
\begin{multline}
\label{SRIVbr}
S\chi_{h^{R}_-,h^{R}_+}^{R,IV,\hslck}(\sigma, \nu, \tau)=
\sum_{a=0}^{u-1} e^{i\pi(u-1-a-m)} \chi^{\isok}_{u-1-a,m}(\sigma,\tau) \\
\times \sum _{b=0}^{u-2}
\vartheta_{(u-1)(m-2m')+u(u-1)(a+1)+2au(\frac{u}{2}-[\frac{u}{2}])-2ub,u(u-1)} \left(\frac{\nu}{u}, \tau\right)\\
\times c^{(u-1)}_{a, a(u-1)+2a(\frac{u}{2}-[\frac{u}{2}])-2b}(\tau)
\end{multline}
and
\begin{multline}
\label{SRVbr}
S\chi_{h^{R}_-,h^{R}_+}^{R,V,\hslck}(\sigma, \nu, \tau)=
\sum_{a=0}^{u-1} e^{i\pi (M+M'-a)} \chi^{\isok}_{u-1-a,u-M-M'-2}(\sigma,\tau) \\
\times \sum _{b=0}^{u-2}
\vartheta_{(u-1)(M'-M)+u(u-1)a+2au(\frac{u}{2}-[\frac{u}{2}])-2ub,u(u-1)} \left(\frac{\nu}{u}, \tau\right)\\
\times c^{(u-1)}_{a, a(u-1)+2a(\frac{u}{2}-[\frac{u}{2}])-2b}(\tau).
\end{multline}

On applying the modular $\Stran$ transformation to a particular character, we obtain a linear combination of class $IV$ and class $V$ characters.  The calculation of the effect of $\Stran$ in the general case is thus simplified by combining the branched $\hslck$ formulae.  (We also recall here that $\Stran$ applied to Neveu-Schwarz characters gives a linear combination of Neveu-Schwarz characters, whereas $\Stran$ applied to Ramond characters gives Neveu-Schwarz supercharacters and vice versa; Ramond supercharacters transform into themselves.)  Examining the Neveu-Schwarz branching formulae \eqref{NSIVbr} and \eqref{NSVbr}, we see that the substitution $M=m'-m-1$, $M'=u-1-m'$ in the class $V$ formula \eqref{NSVbr} gives us precisely the class $IV$ formula \eqref{NSIVbr}.  Since $0\leqslant M+M' \leqslant u-2$, these new values of $m=u-2-M-M'$ and $m'=u-1-M'$ are those for which $0\leqslant m <m' \leqslant u-1$, whereas for class $IV$ we have $0 \leqslant m' \leqslant m \leqslant u-1$.  We can use the same branching formula \eqref{NSIVbr} for both class $IV$ and class $V$, with $0\leqslant m \leqslant u-1$ and $0\leqslant m' \leqslant u-1$, but we must still use the relevant $M$ and $M'$ and expressions for the class $V$ quantum numbers to obtain the correct values of $h^{NS}_-$ and $h^{NS}_+$.

The story for the Ramond characters \eqref{RIVbr} and \eqref{RVbr} is exactly the same; for the Neveu-Schwarz and Ramond supercharacters we must modify \eqref{SNSIVbr} and \eqref{SRIVbr} slightly.  The final versions of the branching formulae read (now labelling the $\hslck$ characters by $m$ and $m'$):
\begin{multline}
\label{NSbr}
\chi_{m,m'}^{NS,\hslck}(\sigma, \nu, \tau)=
\sum_{a=0}^{u-1}  \chi^{\isok}_{a,u-m-1}(\sigma,\tau)\\
\times \sum _{b=0}^{u-2} 
\vartheta_{(u-1)(m-2m')+u(u-1)(a+1)+2au(\frac{u}{2}-[\frac{u}{2}])-2ub,u(u-1)} \left(\frac{\nu}{u}, \tau\right)\\
\times c^{(u-1)}_{a, a(u-1)+2a(\frac{u}{2}-[\frac{u}{2}])-2b}(\tau);
\end{multline}
\begin{multline}
\label{Rbr}
\chi_{m,m'}^{R,\hslck}(\sigma, \nu, \tau)=
\sum_{a=0}^{u-1}  \chi^{\isok}_{u-1-a,m}(\sigma,\tau)\\
\times \sum _{b=0}^{u-2} 
\vartheta_{(u-1)(m-2m')+u(u-1)(a+1)+2au(\frac{u}{2}-[\frac{u}{2}])-2ub,u(u-1)} \left(\frac{\nu}{u}, \tau\right) \\
\times c^{(u-1)}_{a, a(u-1)+2a(\frac{u}{2}-[\frac{u}{2}])-2b}(\tau);
\end{multline}
\begin{multline}
\label{SNSbr}
S\chi_{m,m'}^{NS,\hslck}(\sigma, \nu, \tau)=
\sum_{a=0}^{u-1} (-1)^{G+a-(u-m-1)} \chi^{\isok}_{a,u-m-1}(\sigma,\tau) \\
\times \sum _{b=0}^{u-2}
\vartheta_{(u-1)(m-2m')+u(u-1)(a+1)+2au(\frac{u}{2}-[\frac{u}{2}])-2ub,u(u-1)} \left(\frac{\nu}{u}, \tau\right)\\
\times c^{(u-1)}_{a, a(u-1)+2a(\frac{u}{2}-[\frac{u}{2}])-2b}(\tau);
\end{multline}
and
\begin{multline}
\label{SRbr}
S\chi_{m,m'}^{R,\hslck}(\sigma, \nu, \tau)=
\sum_{a=0}^{u-1}  (-1)^{G+u-1-a-m} \chi^{\isok}_{u-1-a,m}(\sigma,\tau)\\
\times \sum _{b=0}^{u-2} 
\vartheta_{(u-1)(m-2m')+u(u-1)(a+1)+2au(\frac{u}{2}-[\frac{u}{2}])-2ub,u(u-1)} \left(\frac{\nu}{u}, \tau\right) \\
\times c^{(u-1)}_{a, a(u-1)+2a(\frac{u}{2}-[\frac{u}{2}])-2b}(\tau),
\end{multline}
where
\begin{equation*}
G=\begin{cases}
0 \qquad \text{if}\quad m\geqslant m',\\ 
1\qquad \text{if} \quad m < m'
\end{cases}
\end{equation*}
and $0\leqslant m, m' \leqslant u-1$ in both sectors.

\section{Modular $\Stran$ Transformation of $\hslck$ Characters}

The action of $\Stran : (\sigma, \nu, \tau) \rightarrow \left(\frac{\sigma}{\tau}, \frac{\nu}{\tau}, -\frac{1}{\tau}\right)$ on the branched $\hslck$ characters \eqref{NSbr}, \eqref{Rbr}, \eqref{SNSbr} and \eqref{SRbr} may now be obtained by use of \eqref{isotr}, \eqref{thetatr} and \eqref{ctr}.  For example, in the case of the Neveu-Schwarz characters \eqref{NSbr} we find
\begin{multline}
\label{NStrl}
\chi_{m,m'}^{NS,\hslck}\left(\frac{\sigma}{\tau},\frac{\nu}{\tau}, -\frac{1}{\tau} \right)= \frac{e^{i\pi(u-1)\left(\sigma^2-\nu^2\right)/u\tau}}{(u-1)\sqrt{2u(u+1)}}
\sum_{a=0}^{u-1} \sum_{n=0}^{u-1} \sum_{n'=0}^{u-1} S_{a(u-m-1),nn'} \chi^{\isok}_{n,n'}(\sigma,\tau)\\
\times \sum _{b=0}^{u-2} \sum_{r=0}^{2u(u-1)-1} e^{-i\pi r ((u-1)(m-2m')+u(u-1)(a+1)+2au(\frac{u}{2}-[\frac{u}{2}])-2ub)/(u(u-1))}
\vartheta_{r,u(u-1)} \left(\frac{\nu}{u}, \tau \right)\\
\times \underset{a'\equiv b' \; \text{mod }2}{\sum^{u-1}_{a'=0}\,\sum^{u-1}_{b'=-u+2}}
s\left(a,l,a',b'\right) c^{(u-1)}_{a',b'}(\tau),
\end{multline}
where $l=a(u-1)+2a\left(\frac{u}{2}-[\frac{u}{2}]\right)-2b$.

At first sight the problem of extracting from this expression a linear combination of Neveu-Schwarz characters \eqref{NSbr} would appear a fairly challenging task.  In fact we do not need to proceed in this way.  Looking at \eqref{NSbr}, we see that taking only the $a=b=0$ term (say) provides us with a unique linear ``signature'' term for each Neveu-Schwarz character, the coefficient of which is necessarily the coefficient of that particular character.  This signature takes the following form:
\begin{equation}
\label{NSbrs}
\chi_{n,n'}^{NS,\hslck}(\sigma, \nu, \tau) \sim
\chi^{\isok}_{0,u-n-1}(\sigma,\tau) \vartheta_{(u-1)(n-2n'+u),u(u-1)} \left(\frac{\nu}{u}, \tau \right) c^{(u-1)}_{0,0}(\tau).
\end{equation}
The problem of computing the coefficients of $\hslck$ Neveu-Schwarz characters in \eqref{NStrl} is thus reduced to a simple matter of extracting the correct coefficients for the signatures \eqref{NSbrs}.  Hence
\begin{equation}
\label{NStr}
\chi_{m,m'}^{NS,\hslck}\left(\frac{\sigma}{\tau},\frac{\nu}{\tau}, -\frac{1}{\tau} \right)=  e^{i\pi(u-1)(\sigma^2-\nu^2)/u\tau} \sum_{n=0}^{u-1} \sum_{n'=0}^{u-1} S^{NS}_{mm',nn'} \chi_{n,n'}^{NS,\hslck}(\sigma,\nu,\tau)
\end{equation}
where
\begin{multline}
S^{NS}_{mm',nn'}= \frac{1}{u(u-1)} \sqrt{\frac{u}{2(u+1)}} 
\sum_{a=0}^{u-1} \sum_{b=0}^{u-2} S_{a(u-m-1),0(u-n-1)}\\ 
\times e^{-i\pi (n-2n'+u)((u-1)(m-2m')+u(u-1)(a+1)+2au(\frac{u}{2}-[\frac{u}{2}])-2ub)/u}\\
\times \{s(a,l,0,0)+s(a,l,u-1,u-1)\},
\end{multline}
noting that by \eqref{csym} $c^{(u-1)}_{u-1,u-1}(\tau)=c^{(u-1)}_{0,0}(\tau)$.  Expanding the various factors gives
\begin{multline}
\label{S(NS)l}
S^{NS}_{mm',nn'}= \frac{1}{u(u-1)(u+1)}  
\sum_{a=0}^{u-1} \sum_{b=0}^{u-2} (-1)^{(u-m-1)+(a+1)(u-n-1)}\\
\times e^{-i\pi (u+1)(u-m-1)(u-n-1)/u} \sin \left(\frac{\pi (a+1)u}{u+1} \right) \\ 
\times e^{-i\pi (n-2n'+u)((u-1)(m-2m')+u(u-1)(a+1)+2au(\frac{u}{2}-[\frac{u}{2}])-2ub)/u}\\
\times \left\{ \sin \left(\frac{\pi (a+1)}{u+1} \right)+e^{i\pi l} \sin \left(\frac{\pi (a+1)u}{u+1} \right)\right\}
\end{multline}
with as before $l=a(u-1)+2a\left(\frac{u}{2}-[\frac{u}{2}]\right)-2b$.

However, this expression simplifies considerably.  The sum over $b$ can trivially be performed to give a factor $u-1$.  The sum over $a$ can also straightforwardly be found to give a factor $u+1$.  Hence the matrix entries $S^{NS}_{mm',nn'}$ in \eqref{NStr} are given by 
\begin{equation}
S^{NS}_{mm',nn'}=\frac{1}{u} (-1)^{m+n}e^{-i\pi(u+1)(u-m-1)(u-n-1)/u} e^{-i\pi(u-1)(m-2m'+u)(n-2n'+u)/u},
\end{equation} 
a perhaps unexpectedly elegant expression.  

For the Ramond characters and Neveu-Schwarz and Ramond supercharacters we proceed in an essentially similar way: the signature for the Ramond characters is
\begin{equation}
\label{Rbrs}
\chi_{n,n'}^{R,\hslck}(\sigma, \nu, \tau) \sim
\chi^{\isok}_{u-1,n}(\sigma,\tau) \vartheta_{(u-1)(n-2n'+u),u(u-1)} \left(\frac{\nu}{u}, \tau \right) c^{(u-1)}_{0,0}(\tau);
\end{equation}
for the Neveu-Schwarz supercharacters it is
\begin{multline}
\label{SNSbrs}
S\chi_{n,n'}^{NS,\hslck}(\sigma, \nu, \tau) \sim
(-1)^{G-(u-n-1)} \chi^{\isok}_{0,u-n-1}(\sigma,\tau)\\
\times \vartheta_{(u-1)(n-2n'+u),u(u-1)} \left(\frac{\nu}{u}, \tau \right) c^{(u-1)}_{0,0}(\tau);
\end{multline}
and for the Ramond supercharacters it is
\begin{equation}
\label{SRbrs}
\chi_{n,n'}^{R,\hslck}(\sigma, \nu, \tau) \sim
(-1)^{G+u-1-n} \chi^{\isok}_{u-1,n}(\sigma,\tau) \vartheta_{(u-1)(n-2n'+u),u(u-1)} \left(\frac{\nu}{u}, \tau \right) c^{(u-1)}_{0,0}(\tau),
\end{equation}
with $G$ defined as before.  As $\Stran$ transforms Ramond characters into Neveu-Schwarz supercharacters and vice versa we extract the coefficient of \eqref{Rbrs} in the $\Stran$-transformed Neveu-Schwarz supercharacter \eqref{SNSbr} and the coefficient of \eqref{SNSbrs} in the $\Stran$-transformed Ramond character \eqref{Rbr}.  Again, the expressions simplify along similar lines to the Neveu-Schwarz case.  We find
\begin{equation}
\label{Rtr}
\chi_{m,m'}^{R,\hslck}\left(\frac{\sigma}{\tau},\frac{\nu}{\tau},-\frac{1}{\tau} \right)=  e^{i\pi(u-1)(\sigma^2-\nu^2)/u\tau} \sum_{n=0}^{u-1} \sum_{n'=0}^{u-1} S^R_{mm',nn'} S\chi_{n,n'}^{NS,\hslck}(\sigma,\nu,\tau)
\end{equation}
where
\begin{equation}
S^R_{mm',nn'}=\frac{1}{u} (-1)^{G'+m+n+u(u-n-1)}e^{-i\pi(u+1)m(u-n-1)/u} e^{-i\pi(u-1)(m-2m'+u)(n-2n'+u)/u};
\end{equation}
\begin{equation}
\label{SNStr}
S\chi_{m,m'}^{NS,\hslck}\left(\frac{\sigma}{\tau},\frac{\nu}{\tau},-\frac{1}{\tau} \right)=  e^{i\pi(u-1)(\sigma^2-\nu^2)/u\tau} \sum_{n=0}^{u-1} \sum_{n'=0}^{u-1} S^{SNS}_{mm',nn'} \chi_{n,n'}^{R,\hslck}(\sigma,\nu,\tau)
\end{equation}
where
\begin{equation}
S^{SNS}_{mm',nn'}=\frac{1}{u} (-1)^{G+m+n+u(u-m-1)}e^{-i\pi(u+1)(u-m-1)n/u} e^{-i\pi(u-1)(m-2m'+u)(n-2n'+u)/u};
\end{equation}
and
\begin{equation}
\label{SRtr}
S\chi_{m,m'}^{R,\hslck}\left(\frac{\sigma}{\tau},\frac{\nu}{\tau},-\frac{1}{\tau} \right)=  e^{i\pi(u-1)(\sigma^2-\nu^2)/u\tau} \sum_{n=0}^{u-1} \sum_{n'=0}^{u-1} S^{SR}_{mm',nn'} S\chi_{n,n'}^{R,\hslck}(\sigma,\nu,\tau)
\end{equation}
where
\begin{equation}
S^{SR}_{mm',nn'}=\frac{1}{u} (-1)^{G+G'+(u-1)(m+n)}e^{-i\pi(u+1)mn/u} e^{-i\pi(u-1)(m-2m'+u)(n-2n'+u)/u},
\end{equation}
with
\begin{equation*}
G=\begin{cases}
0 \qquad \text{if}\quad m\geqslant m',\\ 
1\qquad \text{if} \quad m < m'
\end{cases}
\end{equation*}
and
\begin{equation*}
G'=\begin{cases}
0 \qquad \text{if}\quad n\geqslant n',\\ 
1\qquad \text{if} \quad n < n'.
\end{cases}
\end{equation*}
We note that $S^{NS}$ and $S^{SR}$ are symmetric; that $S^R = (S^{SNS})^{\textsf{T}}$; that all of these matrices are unitary; and that the matrices as calculated by brute force for $u=2$ (as found in \cite{BHT98}) and $u=3$ (also calculated in \cite{H} for the Neveu-Schwarz characters) given in the appendices agree with the above results. 

In order to consider modular invariant combinations of $\hslck$ characters, we must also know how they transform under the modular $\Ttran$ transformation $\Ttran: (\sigma, \nu, \tau) \rightarrow (\sigma, \nu, \tau + 1)$.  It can be shown that the action of $\Ttran$ is as follows \cite{BHT98, H}:
\begin{align}
\label{ttr}
\chi^{R,\hslck}_{m,m'}(\sigma,\nu,\tau+1)&=e^{2\pi ih^R}\chi^{R,\hslck}_{m,m'}(\sigma,\nu,\tau), \notag\\
\chi^{NS,\hslck}_{m,m'}(\sigma,\nu,\tau+1)&=e^{2\pi ih^{NS}}S\chi^{NS,\hslck}_{m,m'}(\sigma,\nu,\tau), \notag \\
S\chi^{NS,\hslck}_{m,m'}(\sigma,\nu,\tau+1)&=e^{2\pi ih^{NS}}\chi^{NS,\hslck}_{m,m'}(\sigma,\nu,\tau), \notag \\
S\chi^{R,\hslck}_{m,m'}(\sigma,\nu,\tau+1)&=e^{2\pi ih^R}S\chi^{R,\hslck}_{m,m'}(\sigma,\nu,\tau),
\end{align}
recalling that for class $V$ characters ($m<m'$) we must use the appropriate $M$ and $M'$ and the class $V$ formulae to calculate the conformal weights.

\section{Modular Invariants}

With the behaviour of $\hslck$ characters under the modular $\Stran$ and
$\Ttran$ transformations now established, we proceed by looking for modular
invariant combinations of characters.  These could be taken as starting
points in the building of partition functions for rational conformal field
theories.  The canonical example of this is of course the classification of
$\su$ modular invariants, implying the classification of the minimal models,
by Cappelli, Itzykson and Zuber \cite{CIZ} (see also \cite{GAN}), with
minimal superconformal models also considered in \cite{Ca}.  It was found that these partition functions fell into an A-D-E pattern.  This work was extended to the case of admissible $\hat{sl}(2)$ representations by Koh and Sorba \cite{KS} and Lu \cite{Lu}, who obtained a complete classification of modular invariants.

Although we do not attempt to obtain a full classification of fractional level $\hslck$ modular invariants here, we do find all invariants for the cases $u=2$ and $u=3$, special cases of which are analogous to the $A$- and $D$-series obtained in the $\su$ case.  It is straightforward to show that such modular invariants exist for all $u \geqslant 2$.

Modular invariant combinations of characters take the form
\begin{multline}
Z=\sum_{m,m',n,n'=0}^{u-1} N^R_{mm',nn'} \chi^R_{m,m'} \overline{\chi^{R}}_{n,n'} +  N^{NS}_{mm,nn'} \chi^{NS}_{m,m'} \overline{\chi^{NS}}_{n,n'}\\ +  N^{SNS}_{mm',nn'} S\chi^{NS}_{m,m'} \overline{S\chi^{NS}}_{n,n'} + \sum_{a,a',b,b'=0}^{u-1} N^{SR}_{aa',bb'} S\chi^R_{a,a'} \overline{S\chi^R}_{b,b'},
\end{multline}
written in this way to emphasise the fact that the Ramond supercharacters
form a closed set under modular transformations, whereas the remaining
sectors mix as detailed previously.  For ``physical'' modular invariants, the $N_{mm',nn'}$ must be non-negative integers.  In addition,  the identity should be unique so $N^R_{00,00}$ must be equal to 1 (the identity character in this context is $\chi^R_{0,0}$).

For the case of $u=2$, we find two possibilities (see appendix):
\begin{equation}
\label{N2i}
(i) \quad N^R=\begin{pmatrix}
1   & 0  & 0   & 0\\
0   &a   &a-1  & 0\\
0   &a-1 &a    & 0\\
0   & 0  & 0   &1
\end{pmatrix},\quad 
N^{NS}=\begin{pmatrix}
a & 0 & 0 & a-1 \\
0 & 1 & 0 & 0   \\
0 & 0 & 1 & 0   \\
a-1& 0 & 0 & a
\end{pmatrix}   
\end{equation}
or
\begin{equation}
\label{N2ii}
(ii) \quad N^R=\begin{pmatrix}
1   & 0  & 0   & 0\\
0   &a-1   &a  & 0\\
0   &a   &a-1  & 0\\
0   & 0  & 0   &1
\end{pmatrix},\quad 
N^{NS}=\begin{pmatrix}
a & 0 & 0 & a-1 \\
0 & 0 & 1 & 0   \\
0 & 1 & 0 & 0   \\
a-1& 0 & 0 & a
\end{pmatrix}.   
\end{equation}
Clearly for non-negative integer entries, $a \in \N$. There are thus an infinite number of modular invariants, a phenomenon also observed in \cite{Lu}.  For $u=3$ we find a similar situation, with an additional parameter:
\begin{align}
\label{N3i}
(i)\quad N^R &= \begin{pmatrix}
1 & 0 & 0 & 0 & 0 & 0 & 0 & 0 & 0\\
0 & a & b & b & 0 &a-1& 0 & 0 & 0\\
0 & b & a &a-1& 0 & b & 0 & 0 & 0\\
0 & b &a-1& a & 0 & b & 0 & 0 & 0\\
0 & 0 & 0 & 0 &a+b& 0 &a+b-1 & 0 & 0\\
0 &a-1& b & b & 0 & a & 0 & 0 & 0\\
0 & 0 & 0 & 0 &a+b-1 & 0 &a+b& 0 & 0\\
0 & 0 & 0 & 0 & 0 & 0 & 0 & 1 & 0\\
0 & 0 & 0 & 0 & 0 & 0 & 0 & 0 & 1
\end{pmatrix},\nonumber\\
 N^{NS} &= \begin{pmatrix}
a+b& 0 & 0 & 0 & 0 & 0 &a+b-1 & 0 & 0\\
0 & a & b & 0 & 0 & 0 & 0 &a-1& b\\
0 & b & a & 0 & 0 & 0 & 0 & b &a-1\\
0 & 0 & 0 & 1 & 0 & 0 & 0 & 0 & 0\\
0 & 0 & 0 & 0 & 1 & 0 & 0 & 0 & 0\\
0 & 0 & 0 & 0 & 0 & 1 & 0 & 0 & 0\\
a+b-1& 0 & 0 & 0 & 0 & 0 &a+b& 0 & 0\\
0 &a-1& b & 0 & 0 & 0 & 0 & a & b\\
0 & b &a-1& 0 & 0 & 0 & 0 & b & a
\end{pmatrix}
\end{align}
or
\begin{align}
\label{N3ii}
(ii)\quad N^R &= \begin{pmatrix}
1 & 0 & 0 & 0 & 0 & 0 & 0 & 0 & 0\\
0 & b & a &a-1& 0 & b & 0 & 0 & 0\\
0 & a & b & b & 0 &a-1& 0 & 0 & 0\\
0 &a-1& b & b & 0 & a & 0 & 0 & 0\\
0 & 0 & 0 & 0 &a+b& 0 &a+b-1 & 0 & 0\\
0 & b &a-1& a & 0 & b & 0 & 0 & 0\\
0 & 0 & 0 & 0 &a+b-1 & 0 &a+b& 0 & 0\\
0 & 0 & 0 & 0 & 0 & 0 & 0 & 0 & 1\\
0 & 0 & 0 & 0 & 0 & 0 & 0 & 1 & 0
\end{pmatrix},\nonumber\\
 N^{NS} &= \begin{pmatrix}
a+b& 0 & 0 & 0 & 0 & 0 &a+b-1 & 0 & 0\\
0 & b & a & 0 & 0 & 0 & 0 & b &a-1\\
0 & a & b & 0 & 0 & 0 & 0 &a-1& b\\
0 & 0 & 0 & 0 & 0 & 1 & 0 & 0 & 0\\
0 & 0 & 0 & 0 & 1 & 0 & 0 & 0 & 0\\
0 & 0 & 0 & 1 & 0 & 0 & 0 & 0 & 0\\
a+b-1& 0 & 0 & 0 & 0 & 0 &a+b& 0 & 0\\
0 & b &a-1& 0 & 0 & 0 & 0 & b & a\\
0 &a-1& b & 0 & 0 & 0 & 0 & a & b
\end{pmatrix}.
\end{align}
For non-negative integer entries we require $a \in \N$, $b \in \Z_+$.  The ordering of terms in these matrices is as given in the tables A.1, A.2, B.1 and B.2.  (Note that as $\Ttran$ interpolates between Neveu-Schwarz characters and supercharacters, $N^{NS}=N^{SNS}$.)  

In the cases \eqref{N2i} and \eqref{N3i}, when $a=1$ and $b=0$, we find that
$N^R=N^{NS}=I$.  This diagonal invariant we find at all levels, since the
matrices $S$ and $T$ are unitary.  With $a=1$ and $b=0$ in \eqref{N2ii} and
\eqref{N3ii}, the resulting expressions are permutation invariants of the form
\begin{equation}
\sum \chi_{m,m'} \overline{\chi}_{\Pi (m,m')},
\end{equation}
where 
\begin{equation}
\Pi (m,m')=(m,(m-m') \mod{u}).
\end{equation}
Alternatively, labelling the characters by $(h_-,h_+)$ this is equivalent (see earlier definitions) to  $\Pi (h_-,h_+)=(h_-,-h_+)$.  We also see this pattern appearing in the Ramond supercharacters (analysis of which we leave to the appendices).  The question arises as to whether such a permutation invariant can be found at other levels: in fact it is fairly easy to check that such an invariant exists at all levels $k=1/u -1$.  These two situations are analogues of the $A$- and $D$-series invariants of the $\su$ case.  

We might expect the $D$-type invariants to be related to an automorphism of
the fusion rules, in the manner of Schellekens and Yankielowicz \cite{SY}.
This might additionally have some description in terms of automorphisms of
the $\hslc$ Dynkin diagram \cite{diF}: however, as we are dealing with a
superalgebra, the Dynkin diagram is not unique.  In fact, the general
question of fusion rules is not a trivial one for the situation of fractional
level ({\em i.e.} non-unitary) theories.  Both of these issues deserve continued attention.

\section{Conclusion}

We have found expressions for the modular $\Stran$ transformation of $\hslck$
characters at fractional level $k=1/u-1$.  This has allowed us to calculate
all modular invariants for the cases $u=2$ and $u=3$, leading to the
discovery of an $A$-series and $D$-series of modular invariants.  The
derivation of the general $\Stran$ transformation of $\hslck$ characters
neatly rounds off the work of \cite{HT98} and enables us to look at modular
invariants in this framework of affine superalgebras at fractional level, a
subject little studied.  It would certainly be interesting to have a complete
classification of these invariants, generally a non-trivial problem; however,
it would not be unlikely that an A-D-E type classification along the lines of \cite{Lu} might appear, given the underlying presence of $\hat{sl}(2)$.  We make no claims that these invariants constitute fully-fledged partition functions, given the complications entering at fractional level: in these situations considering fusion rules requires the inclusion of fields not corresponding to highest or lowest weight representations, as originally discovered for fractional level $\hat{sl}(2)$ by Awata and Yamada \cite{AY}. Work has also been done on the case of fractional level $\hat{sl}(3)$---see for example \cite{GPW}, which
also contains some discussion of the $\hat{sl}(2)$ case.  As to superalgebras, the fusion rules of admissible representations of $\widehat{osp}(1|2)$ have been studied in \cite{ER}.  It would be interesting to consider fusion rules in the present context and attempt to realise a rational conformal field theory based on fractional level $\hslck$.

\subsection*{Acknowledgements}

GBJ thanks Peter Bowcock and Anne Taormina for useful comments and discussions, and acknowledges the award of an EPSRC research studentship.

\section*{Appendix A: $u=2$}

\def\theequation{\thesection.\arabic{equation}}
\renewcommand{\theequation}{A.\arabic{equation}}
\setcounter{equation}{0}

Here we list the explicit forms of the matrices $S$ for each sector at $u=2$.  We then list the possible modular invariants satisfying the condition that the matrices $N$ have non-negative integer entries.  In what follows, we understand a sum over the repeated index $\beta$.
\begin{multline}
\chi^{NS,\hslcp}_{\alpha}\left(\frac{\sigma}{\tau}, \frac{\nu}{\tau}, -\frac{1}{\tau}\right) = e^{\pi i \left(\sigma^2 - \nu^2\right)/2\tau}\,S^{NS}_{\alpha \beta} \chi^{NS,\hslcp}_{\beta}(\sigma, \nu, \tau),\\
\alpha, \beta = 1, 2, 3, 4,
\end{multline}
where
\begin{equation}
S^{NS}_{\alpha \beta}= \hf
\begin{pmatrix} i & 1 & 1 & i \\
                 1 &-i & i &-1 \\
                 1 & i &-i &-1 \\
                 i &-1 &-1 & i \end{pmatrix};
\end{equation}
\begin{multline}
\chi^{R,\hslcp}_{\alpha}\left(\frac{\sigma}{\tau}, \frac{\nu}{\tau}, -\frac{1}{\tau}\right) = e^{\pi i \left(\sigma^2 - \nu^2\right)/2\tau}\,S^R_{\alpha \beta} S\chi^{NS,\hslcp}_{\beta}(\sigma, \nu, \tau),\\
\alpha, \beta = 1, 2, 3, 4,
\end{multline}
where
\begin{equation}
S^R_{\alpha \beta}= \hf
\begin{pmatrix} 1 & 1 & 1 &-1 \\
                 i &-i & i & i \\
                 i & i &-i & i \\
                 1 &-1 &-1 &-1 \end{pmatrix};
\end{equation}
\begin{multline}
S\chi^{NS,\hslcp}_{\alpha}\left(\frac{\sigma}{\tau}, \frac{\nu}{\tau}, -\frac{1}{\tau}\right) = e^{\pi i \left(\sigma^2 - \nu^2\right)/2\tau}\,S^{SNS}_{\alpha \beta} \chi^{R,\hslcp}_{\beta}(\sigma, \nu, \tau),\\
\alpha, \beta = 1, 2, 3, 4,
\end{multline}
where
\begin{equation}
S^{SNS}_{\alpha \beta}= \hf
\begin{pmatrix} 1 & i & i & 1 \\
                 1 &-i & i &-1 \\
                 1 & i &-i &-1 \\
                -1 & i & i &-1 \end{pmatrix};
\end{equation}
\begin{multline}
S\chi^{R,\hslcp}_{\alpha}\left(\frac{\sigma}{\tau}, \frac{\nu}{\tau}, -\frac{1}{\tau}\right) = e^{\pi i \left(\sigma^2 - \nu^2\right)/2\tau}\,S^{SR}_{\alpha \beta} S\chi^{R,\hslcp}_{\beta}(\sigma, \nu, \tau),\\
\alpha, \beta = 1, 2, 3, 4,
\end{multline}
where
\begin{equation}
S^{SR}_{\alpha \beta}= \hf
\begin{pmatrix} 1 & 1 & 1 & -1 \\
                 1 &1  & -1 &1 \\
                 1 &-1 & 1 & 1 \\
                 -1 &1 &1 & 1 \end{pmatrix}.
\end{equation}

We use the definitions as laid out in the following tables.\\
\\
\centerline{{\footnotesize{Table A.1: class $IV$ $\hslcp$ characters}}}\\
\renewcommand{\arraystretch}{1.5}
\[\begin{array}{|c||c|c||c|c|c||c|c|c|}
\hline
       & m & m' & h^R_{-} & h^R_{+} & h^R & h^{NS}_{-} & h^{NS}_{+} & h^{NS} \\
\hline
\chi_1 & 0 & 0  & 0       & 0       & 0  & -\hf       & 0         & -\egt  \\
\chi_2 & 1 & 0  & -\hf    & -\hf    &0 & 0           & -\hf        & \egt   \\
\chi_3 & 1 & 1  & -\hf    & \hf     &0 & 0           & \hf         & \egt   \\
\hline
\end{array}\]\\
\\
\centerline{{\footnotesize{Table A.2: class $V$ $\hslcp$ characters}}}\\
\[\begin{array}{|c||c|c||c|c|c||c|c|c|}
\hline
       & M(m) & M'(m') & h^R_{-} & h^R_{+} & h^R& h^{NS}_{-} & h^{NS}_{+}  & h^{NS} \\
\hline
\chi_4 & 0(0) & 0(1)  & 1       & 0       & \hf & -\thf      & 0           & -\egt  \\
\hline
\end{array}\]
\\
\\
The supercharacters in each sector have the same quantum numbers as the corresponding characters.  The relation between $M$ and $M'$ values in class $V$ and the $m$ and $m'$ values which allow us to combine classes $IV$ and $V$ in the branching formulae \eqref{NSbr}, \eqref{Rbr}, \eqref{SNSbr} and \eqref{SRbr} is $m=u-2-M-M'$, $m'=u-1-M'$.

With the above information, we have calculated modular invariant matrices $N$ in 
\begin{multline}
Z=\sum_{m,m',n,n'=0}^{u-1} N^R_{mm',nn'} \chi^R_{m,m'} \overline{\chi^{R}}_{n,n'} +  N^{NS}_{mm,nn'} \chi^{NS}_{m,m'} \overline{\chi^{NS}}_{n,n'}\\ +  N^{SNS}_{mm',nn'} S\chi^{NS}_{m,m'} \overline{S\chi^{NS}}_{n,n'} + \sum_{a,a',b,b'=0}^{u-1} N^{SR}_{aa',bb'} S\chi^R_{a,a'} \overline{S\chi^R}_{b,b'},
\end{multline}
that is to say, $N$ such that $[S,N]=[T,N]=0$, using the appropriate matrices $S$ and $T$.  We find that the general form of these $N$ is
\begin{equation*}
N^R=\begin{pmatrix}
a-b & 0 & 0 & 0\\
0   &c+b&a-c& 0\\
0   &a-c&c+b& 0\\
0   & 0 & 0 &a-b
\end{pmatrix},
\end{equation*}
\begin{equation}
N^{NS}=N^{SNS}=\begin{pmatrix}
a & 0 & 0 & b\\
0   &c &a-b-c& 0\\
0   &a-b-c&c & 0\\
b   & 0 & 0 &a
\end{pmatrix}.
\end{equation}

With the requirements that all the $N_{mm',nn'}$ are non-negative integers and $N^R_{00,00}=1$, we find two possible cases:
\begin{equation}
(i) \quad N^R=\begin{pmatrix}
1   & 0  & 0   & 0\\
0   &a   &a-1  & 0\\
0   &a-1 &a    & 0\\
0   & 0  & 0   &1
\end{pmatrix},\quad 
N^{NS}=\begin{pmatrix}
a & 0 & 0 & a-1 \\
0 & 1 & 0 & 0   \\
0 & 0 & 1 & 0   \\
a-1& 0 & 0 & a
\end{pmatrix}   
\end{equation}
or
\begin{equation}
(ii) \quad N^R=\begin{pmatrix}
1   & 0  & 0   & 0\\
0   &a-1   &a  & 0\\
0   &a   &a-1  & 0\\
0   & 0  & 0   &1
\end{pmatrix},\quad 
N^{NS}=\begin{pmatrix}
a & 0 & 0 & a-1 \\
0 & 0 & 1 & 0   \\
0 & 1 & 0 & 0   \\
a-1& 0 & 0 & a
\end{pmatrix},   
\end{equation}
with $a \in \N$. 
For the Ramond supercharacters we find
\begin{equation}
N^{SR}=\begin{pmatrix}
d+e+f+g-h & d+f-h & e+g-h & 0\\
d+e-h &d & e & 0\\
f+g-h & f&g & 0\\
0   & 0 & 0 &h
\end{pmatrix}.
\end{equation}
Setting $d=g=h=1$, $e=f=0$ gives the identity matrix and $e=f=h=1$, $d=g=0$ gives us the permutation.

\section*{Appendix B: $u=3$}

\renewcommand{\theequation}{B.\arabic{equation}}
\setcounter{equation}{0}

\begin{multline}
\chi^{NS,\hslcq}_{\alpha}\left(\frac{\sigma}{\tau}, \frac{\nu}{\tau}, -\frac{1}{\tau}\right) = e^{2\pi i \left(\sigma^2 - \nu^2\right)/3\tau}\,S^{NS}_{\alpha \beta} \chi^{NS,\hslcq}_{\beta}(\sigma, \nu, \tau),\\
\alpha, \beta = 1, 2, \dots, 9,
\end{multline}
where
\begin{equation}
S^{NS}_{\alpha \beta}= \frac{1}{3}
\begin{pmatrix}
\ruii & \rui  & \rui  & 1     &  1  & 1     & \rui  & \ruii & \ruii \\
\rui  &  1    & \rumii& \rumi & -1  & \rui  & \ruii & \rumi & -1    \\
\rui  & \rumii&   1   & \rui  & -1  & \rumi & \ruii &   -1  & \rumi \\
   1  & \rumi & \rui  & \rumii&  1  & \ruii &   -1  & \ruii & \rumii\\
   1  &  -1   &   -1  &   1   &  1  &   1   &   -1  &   1   &  1    \\
   1  & \rui  & \rumi & \ruii &  1  & \rumii&   -1  & \rumii& \ruii \\
\rui  & \ruii & \ruii &  -1   & -1  & -1    & \ruii & \rui  & \rui  \\
\ruii & \rumi &  -1   & \ruii &  1  & \rumii& \rui  &  1    & \rumii\\
\ruii &  -1   & \rumi & \rumii&  1  & \ruii & \rui  & \rumii&  1    
\end{pmatrix};
\end{equation}
\begin{multline}
\chi^{R,\hslcq}_{\alpha}\left(\frac{\sigma}{\tau}, \frac{\nu}{\tau}, -\frac{1}{\tau}\right) = e^{2\pi i \left(\sigma^2 - \nu^2\right)/3\tau}\,S^R_{\alpha \beta} S\chi^{NS,\hslcq}_{\beta}(\sigma, \nu, \tau),\\
\alpha, \beta = 1, 2, \dots, 9,
\end{multline}
where
\begin{equation}
S^R_{\alpha \beta}= \frac{1}{3}
\begin{pmatrix}
  1   &  1    & 1     & 1     &  1  & 1     & -1    & -1    & -1    \\
\rui  & -1    & \rui  & \rumi & -1  & \rui  & \ruii & \ruii & 1     \\
\rui  & \rui  & -1    & \rui  & -1  & \rumi & \ruii & 1     & \ruii \\
\ruii & 1     & \ruii & \rumii&  1  & \ruii & \rui  & \rui  & -1    \\
\ruii & \rumii& \rumii&   1   &  1  &   1   & \rui  & \rumi & \rumi \\
\ruii & \ruii & 1     & \ruii &  1  & \rumii& \rui  & -1    & \rui  \\
\rui  & \rumi & \rumi &  -1   & -1  & -1    & \ruii & \rumii& \rumii\\
 1    & \rumii& \ruii & \ruii &  1  & \rumii&  -1   & \rui  & \rumi \\
 1    & \ruii & \rumii& \rumii&  1  & \ruii &  -1   & \rumi & \rui    
\end{pmatrix};
\end{equation}
\begin{multline}
S\chi^{NS,\hslcq}_{\alpha}\left(\frac{\sigma}{\tau}, \frac{\nu}{\tau}, -\frac{1}{\tau}\right) = e^{2\pi i \left(\sigma^2 - \nu^2\right)/3\tau}\,S^{SNS}_{\alpha \beta} \chi^{R,\hslcq}_{\beta}(\sigma, \nu, \tau),\\
\alpha, \beta = 1, 2, \dots, 9,
\end{multline}
where
\begin{equation}
S^{SNS}_{\alpha \beta}= \frac{1}{3}
\begin{pmatrix}
  1   & \rui  & \rui  & \ruii &\ruii  & \ruii & \rui  & 1     & 1     \\
  1   & -1    & \rui  & 1     &\rumii & \ruii & \rumi & \rumii& \ruii \\
  1   & \rui  & -1    & \ruii &\rumii & 1     & \rumi & \ruii & \rumii\\
  1   & \rumi &  \rui & \rumii&  1    & \ruii & -1    & \ruii & \rumii\\
  1   &  -1   &  -1   &   1   &  1    &   1   & -1    &  1    &  1    \\
  1   & \rui  & \rumi & \ruii &  1    & \rumii& -1    & \rumii& \ruii \\
 -1   & \ruii & \ruii & \rui  & \rui  & \rui  & \ruii & -1    & -1    \\
 -1   & \ruii & 1     & \rui  & \rumi & -1    & \rumii& \rui  & \rumi \\
 -1   & 1     & \ruii & -1    & \rumi & \rui  & \rumii& \rumi & \rui    
\end{pmatrix};
\end{equation}
\begin{multline}
S\chi^{R,\hslcq}_{\alpha}\left(\frac{\sigma}{\tau}, \frac{\nu}{\tau}, -\frac{1}{\tau}\right) = e^{2\pi i \left(\sigma^2 - \nu^2\right)/3\tau}\,S^{SR}_{\alpha \beta} S\chi^{R,\hslcq}_{\beta}(\sigma, \nu, \tau),\\
\alpha, \beta = 1, 2, \dots, 9,
\end{multline}
where
\begin{equation}
S^{SR}_{\alpha \beta}= \frac{1}{3}
\begin{pmatrix}
  1   & 1     & 1     & 1     &  1    & 1     & -1    & -1    & -1    \\
  1   & 1     & \rumii& 1     &\rumii & \ruii & \rumi & \rui  & \rumi \\
  1   & \rumii& 1     & \ruii &\rumii & 1     & \rumi & \rumi & \rui  \\
  1   & 1     &  \ruii& 1     & \ruii & \rumii& \rui  & \rumi & \rui  \\
  1   & \rumii& \rumii& \ruii & \ruii & \ruii & \rui  & -1    & -1    \\
  1   & \ruii & 1     & \rumii& \ruii & 1     & \rui  & \rui  & \rumi \\
 -1   & \rumi & \rumi & \rui  & \rui  & \rui  & \ruii &  1    & 1     \\
 -1   & \rui  & \rumi & \rumi & -1    & \rui  & 1     & \rumii& \ruii \\
 -1   & \rumi & \rui  & \rui  & -1    & \rumi & 1     & \ruii & \rumii   
\end{pmatrix}.
\end{equation}
\pagebreak
In the above we use the following definitions:\\
\\
\centerline{{\footnotesize{Table B.1: class $IV$ $\hslcq$ characters}}}\\
\renewcommand{\arraystretch}{1.5}
\[\begin{array}{|c||c|c||c|c|c||c|c|c|}
\hline
      & m & m' & h^R_{-} & h^R_{+} & h^R  &h^{NS}_{-} & h^{NS}_{+} & h^{NS} \\
\hline
\chi_1 & 0 & 0 & 0       & 0       &0     & -\tthrd & 0            & -\sxth \\
\chi_2 & 1 & 0 & -\thrd  & -\thrd  &0     & -\thrd  & -\thrd       & 0      \\
\chi_3 & 1 & 1 & -\thrd  & \thrd   &0     & -\thrd  & \thrd        & 0      \\
\chi_4 & 2 & 0 & -\tthrd & -\tthrd &0     & 0       & -\tthrd      & \sxth  \\
\chi_5 & 2 & 1 & -\tthrd & 0       &\thrd & 0       & 0            & \hf    \\
\chi_6 & 2 & 2 & -\tthrd & \tthrd  &0     & 0       & \tthrd       & \sxth  \\
\hline
\end{array}\]\\
\\
\centerline{{\footnotesize{Table B.2: class $V$ $\hslcq$ characters}}}\\
\[\begin{array}{|c||c|c||c|c|c||c|c|c|}
\hline
       & M(m) & M'(m') & h^R_{-} & h^R_{+} &h^R& h^{NS}_{-} & h^{NS}_{+}  & h^{NS} \\
\hline
\chi_7 & 0(1) & 0(2)  & \tthrd  & 0         & \thrd& -\fthrd    & 0       & -\sxth \\
\chi_8 & 0(0) & 1(1)  & 1       & -\thrd   &\tthrd& -\vthrd    & -\thrd   & 0      \\
\chi_9 & 1(0) & 0(2)  & 1       & \thrd   &\tthrd& -\vthrd    & \thrd     & 0      \\
\hline
\end{array}\]

For the $u=3$ modular invariants we find:
\begin{equation*}
N^R=\begin{pmatrix}
a-b+c-d & 0 &0 &0 &0 &0 &0 &0 &0\\
0 & a & c & b &0 &d &0 &0 &0 \\
0 & c & a & d &0 & b & 0& 0& 0\\
0 & b & d & a & 0 & c& 0&0&0\\
0&0&0&0&a+c &0&b+d &0&0\\
0& d & b & c & 0& a&0&0&0\\
0&0&0&0&b+d&0&a+c&0&0\\
0&0&0&0&0&0&0&a-d&c-b\\
0&0&0&0&0&0&0&c-b&a-d
\end{pmatrix},
\end{equation*}
\begin{equation}
N^{NS}=N^{SNS}=\begin{pmatrix}
a+c & 0 &0 &0 &0 &0 &b+d &0 &0\\
0 & a & c  &0 &0 &0 &0 &d &b \\
0 & c& a & 0&0 & 0 & 0& b& d\\
0 & 0 & 0 & a-d & 0 & c-b& 0&0&0\\
0&0&0&0&a-b+c-d &0&0 &0&0\\
0& 0 & 0 & c-b& 0& a-d&0&0&0\\
b+d&0&0&0&0&0&a+c&0&0\\
0&d&b&0&0&0&0&a&c\\
0&b&d&0&0&0&0&c&a
\end{pmatrix}.
\end{equation}
Requiring that entries be non-negative integers and that $N^R_{00,00}=1$ leads to the following invariants:
\begin{equation*}
(i)\quad N^R = \begin{pmatrix}
1 & 0 & 0 & 0 & 0 & 0 & 0 & 0 & 0\\
0 & a & b & b & 0 &a-1& 0 & 0 & 0\\
0 & b & a &a-1& 0 & b & 0 & 0 & 0\\
0 & b &a-1& a & 0 & b & 0 & 0 & 0\\
0 & 0 & 0 & 0 &a+b& 0 &a+b-1 & 0 & 0\\
0 &a-1& b & b & 0 & a & 0 & 0 & 0\\
0 & 0 & 0 & 0 &a+b-1 & 0 &a+b& 0 & 0\\
0 & 0 & 0 & 0 & 0 & 0 & 0 & 1 & 0\\
0 & 0 & 0 & 0 & 0 & 0 & 0 & 0 & 1
\end{pmatrix},
\end{equation*}
\begin{equation}
N^{NS} = \begin{pmatrix}
a+b& 0 & 0 & 0 & 0 & 0 &a+b-1 & 0 & 0\\
0 & a & b & 0 & 0 & 0 & 0 &a-1& b\\
0 & b & a & 0 & 0 & 0 & 0 & b &a-1\\
0 & 0 & 0 & 1 & 0 & 0 & 0 & 0 & 0\\
0 & 0 & 0 & 0 & 1 & 0 & 0 & 0 & 0\\
0 & 0 & 0 & 0 & 0 & 1 & 0 & 0 & 0\\
a+b-1& 0 & 0 & 0 & 0 & 0 &a+b& 0 & 0\\
0 &a-1& b & 0 & 0 & 0 & 0 & a & b\\
0 & b &a-1& 0 & 0 & 0 & 0 & b & a
\end{pmatrix}
\end{equation}
or
\begin{equation*}
(ii)\quad N^R = \begin{pmatrix}
1 & 0 & 0 & 0 & 0 & 0 & 0 & 0 & 0\\
0 & a & c &c-1& 0 & a & 0 & 0 & 0\\
0 & c & a & a & 0 &c-1& 0 & 0 & 0\\
0 &c-1& a & a & 0 & c & 0 & 0 & 0\\
0 & 0 & 0 & 0 &a+c& 0 &a+c-1 & 0 & 0\\
0 & a &c-1& c & 0 & a & 0 & 0 & 0\\
0 & 0 & 0 & 0 &a+c-1 & 0 &a+c& 0 & 0\\
0 & 0 & 0 & 0 & 0 & 0 & 0 & 0 & 1\\
0 & 0 & 0 & 0 & 0 & 0 & 0 & 1 & 0
\end{pmatrix},
\end{equation*}
\begin{equation}
N^{NS} = \begin{pmatrix}
a+c& 0 & 0 & 0 & 0 & 0 &a+c-1 & 0 & 0\\
0 & a & c & 0 & 0 & 0 & 0 & a &c-1\\
0 & c & a & 0 & 0 & 0 & 0 &c-1& a\\
0 & 0 & 0 & 0 & 0 & 1 & 0 & 0 & 0\\
0 & 0 & 0 & 0 & 1 & 0 & 0 & 0 & 0\\
0 & 0 & 0 & 1 & 0 & 0 & 0 & 0 & 0\\
a+c-1& 0 & 0 & 0 & 0 & 0 &a+c& 0 & 0\\
0 & a &c-1& 0 & 0 & 0 & 0 & a & c\\
0 &c-1& a & 0 & 0 & 0 & 0 & c & a
\end{pmatrix}.
\end{equation}
For the Ramond supercharacters we find
\begin{equation}
N^{SR}=\begin{pmatrix}
e_1 & e_2 &e_3 &e_2 &0 &e_3 &0 &0 &0\\
g & j &f_3  &f_4 &0 &h &0 &0 &0 \\
e & g_2&g_3 & f&0 & l & 0& 0& 0\\
g & f_4 & h & j & 0 & f_3& 0&0&0\\
0&0&0&0&k &0&j_7 &0&0\\
e& f & l & g_2& 0& g_3&0&0&0\\
0&0&0&0&j_7&0&k&0&0\\
0&0&0&0&0&0&0&m_8&m_9\\
0&0&0&0&0&0&0&m_9&m_8
\end{pmatrix}
\end{equation}
where
$e_1=f+g+k+l$, $e_2=f+g-h$, $e_3=e-f+h$, $f_3=g-j+k$, $f_4=-e+f+g-h+l$, $g_2=f+g-h-j+k$, $g_3=e-f-g+h+j$, $j_7=e-f-l$, $m_8=-g+h+j$, $m_9=-e+f+g-h-j+k+l$.  Setting $j=k=1$, $e=f=g=h=l=0$ gives us the identity and $k=1$, $e=f=g=h=j=l=0$ the permutation invariant.

\def\NPB{Nucl.\ Phys.\ B } 
\def\CMP{Commun.\ Math.\ Phys.\ }

\end{document}